\documentclass[aps,prc,superscriptaddress,twoside,twocolumn,nofootinbib,10pt,%
showpacs,floatfix]{revtex4-1}

\usepackage{amsmath,amssymb}
\usepackage{graphicx,bm}
\usepackage{slashed}
\usepackage{epstopdf}
\usepackage{ulem} %% for strike-through
\usepackage[usenames]{color}
\usepackage{float}

\renewcommand\sout{\bgroup \color{red} \ULdepth=-.5ex \ULset}

\begin{document}
\title{Observation of $T_{cc}$ and a quark model}

\author{Sungsik Noh}\affiliation{Department of Physics and Institute of Physics and Applied Physics, Yonsei University, Seoul 03722, Korea}
\author{Woosung Park}\affiliation{Department of Physics and Institute of Physics and Applied Physics, Yonsei University, Seoul 03722, Korea}
%\date{\today}

\begin{abstract}
The recent discovery of the doubly charmed tetraquark $T_{cc}$ ($\bar{u}\bar{d}cc$) provides a stringent constraint on its binding energy relative to its lowest decay threshold. We use a fully convergent spatial wave function and perform a simultaneous global fit to both the meson and baryon spectra. Our analysis shows that a Yukawa type hyperfine potential leads to a slight bound state for $T_{cc}$ with $(I,S) = (0,1)$ below its lowest threshold, in agreement with recent experimental findings. We also find that $T_{cc}$ is highly likely to be in a compact configuration.
\end{abstract}


\maketitle

%%%%%%%%%%%%%%%%%%%%%%%%%%%%%%%%%%%%%%%%%%%%%%%%%%%%%%%%%%%%%%%%%%%%%%%%%%%%%%%%%%%%%%%%%%%%%%%
%{\it Introduction:}
%%%%%%%%%%%%%%%%%%%%%%%%%%%%%%%%%%%%%%%%%%%%%%%%%%%%%%%%%%%%%%%%%%%%%%%%%%%%%%%%%%%%%%%%%%%%%%%
There is a renewed excitement in hadron physics over the  observations of numerous exotic hadron candidates \cite{LHCb:PRL2015,Fermi:PRL2016,LHCb:PRL2019,Choi:2003ue}. Of particular interest is the recently observed flavor exotic $T_{cc}$ state, which is an explicit four quark state. $T_{cc}$ was first predicted in Refs.~\cite{Ballot:1983iv,Zouzou:1986qh} based on the strong color-spin attraction of $\bar{u}\bar{d}$ quark pair that can bind the tetraquark configuration. Many quark model calculations for $T_{cc}$ followed~\cite{ComRef1,ComRef2,ComRef3,ComRef4,ComRef5,ComRef6,ComRef7,ComRef9,ComRef10,ComRef12,ComRef15,ComRef16,ComRef17,Woosung:NPA2019,ComRef23,ComRef24,ComRef25,ComRef27,ComRef30,Noh:Prd2021,ComRef32,Vijande:2009kj,Tan:2020ldi}, which unfortunately varied on the predicted masses and even on whether the mass lies above or below the lowest threshold. Therefore, the recent observation of $T_{cc}$~\cite{LHCbTcc} provides a good opportunity where all the models can be tested in a hitherto untested multiquark configuration, thereby leading us to identify the correct model to describe the low energy confinement phenomena of quantum chromodynamcis (QCD).

Accurate model calculations are a primary requirement for fully testing a quark model. Thus, one first has to introduce a complete set of spatial wave functions, which necessarily contain all possible internal states. Also, a simultaneous global fit to both the meson and baryon spectra should be performed in the model calculations. However, only a few works \cite{Noh:Prd2021,ComRef3,ComRef9,Vijande:2009kj} satisfy these requirements among the works mentioned above~\cite{Ballot:1983iv,Zouzou:1986qh,ComRef1,ComRef2,ComRef3,ComRef4,ComRef5,ComRef6,ComRef7,ComRef9,ComRef10,ComRef12,ComRef15,ComRef16,ComRef17,Woosung:NPA2019,ComRef23,ComRef24,ComRef25,ComRef27,ComRef30,Noh:Prd2021,ComRef32,Vijande:2009kj,Tan:2020ldi}. Furthermore, no work could correctly predict both the mass and binding energy of $T_{cc}$ simultaneously. For example, in our latest publication~\cite{Noh:Prd2021}, we successfully predicted the mass of $T_{cc}$ using a Gaussian type hyperfine potential, whereas the binding energy was obtained to be higher than the experimental measurement by 13 MeV.

The hyperfine potential strongly affects the binding energy of the multiquark configuration, with the strong color-spin attraction coming from the $\bar{u}\bar{d}$ quark pair. 
We thus analyze the effect from a Yukawa type hyperfine potential on the binding energy in comparison with the Gaussian type hyperfine potential.
Our analysis suggests that a Yukawa type hyperfine potential is necessary to accurately reproduce the experimentally observed slight binding of $T_{cc}$, rather than a Gaussian type.

On the other hand, there are chiral quark models based on color-flavor interaction. However, those model studies predicted bindings that are too strong in the $T_{cc}$ channel~\cite{ComRef4,ComRef6,ComRef9,ComRef12,Vijande:2009kj,Tan:2020ldi}. Thus, any modification should start from the gluon exchange quark models.

%%%%%%%%%%%%%%%%%%%%%%%%%%%%%%%%%%%%%%%%%%%%%%%%%%%%%%%%%%%%%%%%%%%%%%%%%%%%%%%%%%%%%%%%%%%%%%%
{\it Model description:}
%%%%%%%%%%%%%%%%%%%%%%%%%%%%%%%%%%%%%%%%%%%%%%%%%%%%%%%%%%%%%%%%%%%%%%%%%%%%%%%%%%%%%%%%%%%%%%%
In our nonrelativisitc quark model, we solve the Schr$\ddot{\rm o}$dinger equation with the Hamiltonian given as follows.
\begin{eqnarray}
H &=& \sum^{4}_{i=1} \left( m_i+\frac{{\mathbf p}^{2}_i}{2 m_i} \right)-\frac{3}{4}\sum^{4}_{i<j}\frac{\lambda^{c}_{i}}{2} \,\, \frac{\lambda^{c}_{j}}{2} \left( V^{C}_{ij} + V^{CS}_{ij} \right), \qquad
\label{Hamiltonian}
\end{eqnarray}
where the confinement potential $V^C_{ij}$ is identical to that used in previous studies \cite{Noh:Prd2021, ComRef3, ComRef9, Vijande:2009kj}.
However, for the hyperfine potential $V^{CS}_{ij}$, we introduce a Yukawa type potential given by
\begin{eqnarray}
V^{CS}_{ij} &=& \frac{\hbar^2 c^2 \kappa'}{m_i m_j c^4} \frac{e^{- r_{ij} / r_{0ij}}}{(r_{0ij}) r_{ij}} \boldsymbol{\sigma}_i \cdot \boldsymbol{\sigma}_j.
\label{CSP}
\end{eqnarray}
Our Yukawa type potential in Eq.~(\ref{CSP}) satisfies the contact term $\delta(r_{ij})$ in the heavy quark limit as $r_{ij}$ approaches zero. In addition, $\kappa'$ and $r_{0ij}$ depend on masses of a quark pair as given in Ref.~\cite{Noh:Prd2021}. The model parameters in the Hamiltonian in Eq.~(\ref{Hamiltonian}) are determined by fitting them to a total of 33 ground state hadron masses listed in Tables~I, II of Ref.~\cite{Noh:Prd2021}. Further, we obtained an optimized set of model parameters in such a way that $\chi^2$ value of the Pearson's chi-squared test formula should be minimized. The model parameters selected for this study are $\kappa=97.7$ MeVfm, $a_0=0.0327338$ (MeV$^{-1}$fm)$^{1/2}$, $m_{u}=315$ MeV, $m_{s}=610$ MeV, $m_{c}=1895$ MeV, $m_{b}=5274$ MeV, $\alpha=1.1349$ fm$^{-1}$, $\beta=0.0011554$ (MeVfm)$^{-1}$, $\gamma=0.001370$ MeV$^{-1}$, $\kappa_0=213.244$ MeV,  $D=959$ MeV.

With this optimized set of parameters, we find that the masses of mesons which compise the strong deacy thresholds for $T_{QQ'}$ states fit well with the experimentally measured masses as shown in Table~\ref{mesons}.

%%%%%%%%%%%%%%%%%%%%%%%%%%%		Table 1. Meson masses		%%%%%%%%%%%%%%%%%%%%%%%%%%%%%%%%%%%%%%
\begin{table}[h!]

\caption{Masses of the lowest decay threshold mesons for $T_{QQ'}$ states, where $Q$ or $Q'$ is a heavy quark($c$ or $b$). $M^{Exp}$ represents the experimentally measured mass, while $M^{Mod}$ is the mass obtained in this work. All masses are given in MeV.}

\centering

\begin{tabular}{cccccc}
\hline
\hline
				&	$D$			&	$D^*$		&	$B$			&	$B^*$		\\
\hline 
	$M^{Exp}$		&	1864.8		&	2007.0	&	5279.3		&	5325.2		\\
	$M^{Mod}$		&	1865.0		&	2009.4	&	5276.3		&	5331.8	\\
\hline 
\hline
\label{mesons}
\end{tabular}
\end{table}
%%%%%%%%%%%%%%%%%%%%%%%%%%%%%%%%%%%%%%%%%%%%%%%%%%%%%%%%%%%%%%%%%%%%%%%%%%%%%%%%%%%%%%%%%%%%%%%
We utilize the same methods as in Ref.~\cite{Noh:Prd2021} to construct the wave function and calculate the masses of tetraquarks as well as those of mesons and baryons.

%%%%%%%%%%%%%%%%%%%%%%%%%%%%%%%%%%%%%%%%%%%%%%%%%%%%%%%%%%%%%%%%%%%%%%%%%%%%%%%%%%%%%%%%%%%%%%%
{\it Masses and binding energies of tetraquarks:}
%%%%%%%%%%%%%%%%%%%%%%%%%%%%%%%%%%%%%%%%%%%%%%%%%%%%%%%%%%%%%%%%%%%%%%%%%%%%%%%%%%%%%%%%%%%%%%%
The results of our calculations for $T_{cc}(\bar{u}\bar{d}cc)$, $T_{cb}(\bar{u}\bar{d}cb)$, and $T_{bb}(\bar{u}\bar{d}bb)$ using our Yukawa type hyperfine potential are presented in Table~\ref{result}. Comparing the present results for $T_{cc}$ with those from our previous publication~\cite{Noh:Prd2021} which used a Gaussian type hyperfine potential, we find that both models reproduce the mass well. However, our previous calculation predicted an unbound $T_{cc}$ state, while our current model indicates that it lies slightly below the threshold consistent with experimental observations. This suggests that the Yukawa type hyperfine potential used in our current calculation may better capture the strong interaction dynamics of the $T_{cc}$ configuration. Further research is necessary to fully understand the implications of our findings.

Our present model using a Yukawa type hyperfine potential shows significantly stronger binding energies for $T_{cc}$ and $T_{cb}$ than the model with a Gaussian type hyperfine potential presented in Ref.~\cite{Noh:Prd2021}.

The discovery of $T_{cc}$ is of great significance since it allows for testing the validity of quark models. In this regard, we compare our present results with those from Refs.\cite{ComRef3,ComRef9,Vijande:2009kj,Noh:Prd2021} in Table\ref{result}. The results from those quark models were also calculated with a complete set of spatial wave functions, but obtained using different forms of Gaussian or Yukawa type of hyperfine potential. Specifically, Refs.\cite{Noh:Prd2021,ComRef3} used a Gaussian type hyperfine potential, while Refs.\cite{ComRef9,Vijande:2009kj}, including our present work, used a Yukawa type of hyperfine potential.

Comparing our results with those from Refs.\cite{ComRef3,ComRef9,Vijande:2009kj,Noh:Prd2021} in Table\ref{result}, the bound $T_{cc}$ ground state is found only in our present work and Refs.\cite{ComRef9,Vijande:2009kj} where the hyperfine potential is of the Yukawa type. Therefore, one can reach a conclusion that a Yukawa type hyperfine potential is necessary to accurately describe the short range interactions within the tetraquark system. The difference in the binding energy between Refs.~\cite{ComRef9,Vijande:2009kj} and our present work consists largely in the difference in the detailed form of the hyperfine potentials used in each model.

To investigate the effect of the forms of hyperfine potential on the binding energy and size of the $T_{cc}$ configuration, we compare the contributions from the Yukawa and Gaussian types of potentials.

%%%%%%%%%%%%%%%%%%%%%%%%%		Table 2. Tetraquark Results		%%%%%%%%%%%%%%%%%%%%%%%%%%%%%
\begin{widetext}

\begin{table}[h!]
\caption{Masses and binding energies ($B_T$) of tetraquark states. $B_T$ is defined as the difference between the tetraquark mass and the sum of the masses of the lowest threshold mesons in the model calculations: $B_T \equiv M_{\text{Tetraquark}} - M_{\text{meson 1}} - M_{\text{meson 2}}$. The mass and binding energy of $T_{cc}$($\bar{u}\bar{d}cc$) reported by LHCb~\cite{LHCbTcc} are 3875 MeV and -0.273 MeV, respectively. All the values are given in MeV.}
\begin{tabular}{ccccccccccccccccc}
\hline
\hline
			&&			&&				&&	Present work	&&&	Ref.~\cite{ComRef3}	&&&	Refs.~\cite{ComRef9,Vijande:2009kj}   &&&	Ref.~\cite{Noh:Prd2021}	\\
\cline{7-8}\cline{10-11}\cline{13-14}\cline{16-17}
	Type		&&	$I(J^P)$	&&	Thresholds	&&	Mass	&$B_{T}$	&&	Mass		&	$B_{T}$	&&	Mass		&	$B_{T}$	&&	Mass		&	$B_{T}$	\\
\hline 
%%% Tbb %%%%%
$\bar{u}\bar{d}bb$	&&	$0(1^+)$	&&	$BB^*$	&&	10464  		&-144	    &&	10506	&	-142		&&	10507	&	-144	&&	10517	&	-145		\\
%%% Tcc %%%%%
$\bar{u}\bar{d}cc$	&&	$0(1^+)$	&&	$DD^*$	&&	3872	&-2 		&&	3931		&	+11		&&	3899		&	-7	   &&	3873		&	+13		\\
%%% Tcb %%%%%
$\bar{u}\bar{d}cb$	&&	$0(1^+)$	&&	$DB^*$	&&	7179	&-18		&&	7244		&	-5		&&	-	&	-   &&	7212		&	-3		\\
\hline 
\hline
\label{result}
\end{tabular}
\end{table}

\end{widetext}
%%%%%%%%%%%%%%%%%%%%%%%%%%%%%%%%%%%%%%%%%%%%%%%%%%%%%%%%%%%%%%%%%%%%%%%%%%%%%%%%%%%%%%%%%%%%%

%%%%%%%%%%%%%%%%%%%%%%%%%%%%%%%%%%%%%%%%%%%%%%%%%%%%%%%%%%%%%%%%%%%%%%%%%%%%%%%%%%%%%%%%%%%%%%%
{\it Detailed analysis of hyperfine potential:}
%%%%%%%%%%%%%%%%%%%%%%%%%%%%%%%%%%%%%%%%%%%%%%%%%%%%%%%%%%%%%%%%%%%%%%%%%%%%%%%%%%%%%%%%%%%%%%%
In the $T_{cc}$ configuration, an important contribution to the binding comes from the hyperfine potential, which can be isolated as follows:
\begin{eqnarray}
B^{Hyp} \equiv H^{Hyp}_{Tetraquark} - H^{Hyp}_{Meson1} - H^{Hyp}_{Meson2} \, ,
\label{HypBinding}
\end{eqnarray}
where $H^{Hyp}_{Tetraquark} \equiv \sum^4_{i<j} V^{CS}(i,j)$ and $H^{Hyp}_{Meson}$'s are the hyperfine part of the Hamiltonian calculated with the corresponding total wave functions for the tetraquark and the mesons, respectively.

%%%%%%%%%%%%%%%%		Table 3. Hyperfine potential separations for TQQ 		%%%%%%%%%%%%%%%%%%%%%
\begin{table}[t]

\caption{Hyperfine potentials $V^{CS}$(in MeV) and relative length $l$(in fm). For $T_{cc}$ configuration, we label the position of each quark as $\bar{u}(1)\bar{d}(2)c(3)c(4)$. The values for the quark pairs of $(1,4), (2,3),$ and $(2,4)$ are the same as those of the $(1,3)$ pair due to symmetry. The subscripts Y and G represent the results obtained using the hyperfine potential in Eq.~(\ref{CSP}) and the Gaussian potential from Ref.~\cite{Noh:Prd2021}, respectively. The values of $B^{Hyp}$ are calculated by Eq.~(\ref{HypBinding}). The RMS ratio of $T_{cc}$ to its threshold $DD^*$ is presented in column 8.}	

\centering

\begin{tabular}{ccccccccccc}
\hline
\hline
Type	&\multicolumn{3}{c}{$T_{cc}$}&&	\multicolumn{2}{c}{Threshold}&&	$B^{Hyp}$	&&	RMS ratio\\
\cline{2-4}\cline{6-7}
				&(1,2)	&(1,3)	&(3,4)		&&	$D$		&	$D^*$	&&		\\
\hline 
$V^{CS}_{\rm Y}$	&-112	&-16.8	&5.0			&&	-120	&	32.5		&&-87	&&	0.89\\
$l_{\rm Y}$		&0.87	&0.71	&0.64		&&	0.55	&	0.61		&&		\\
\hline
$V^{CS}_{\rm G}$	&-109	&-17.4	&5.3			&&	-127	&	34.1		&&-80	&&	0.87\\
$l_{\rm G}$		&0.83&0.67		&0.61		&&	0.52	&	0.59		&&		\\
\hline 
\hline
\end{tabular}
\label{HypSep}
\end{table}
%%%%%%%%%%%%%%%%%%%%%%%%%%%%%%%%%%%%%%%%%%%%%%%%%%%%%%%%%%%%%%%%%%%%%%%%%%%%%%%%%%%%%%%%%%%%%%%%%%

\begin{figure}[h!]
\centering

\includegraphics[width=0.49\columnwidth]{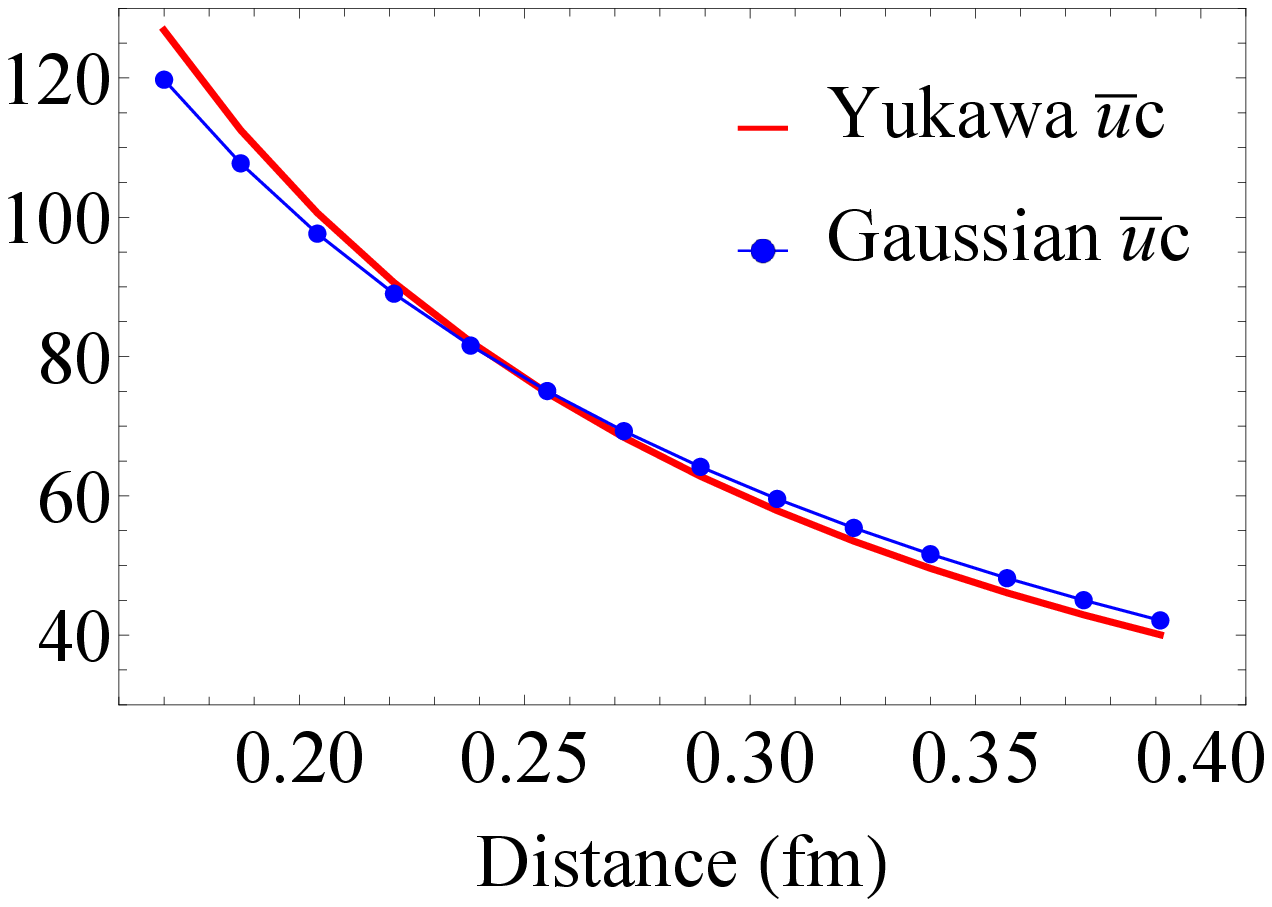}
\includegraphics[width=0.48\columnwidth]{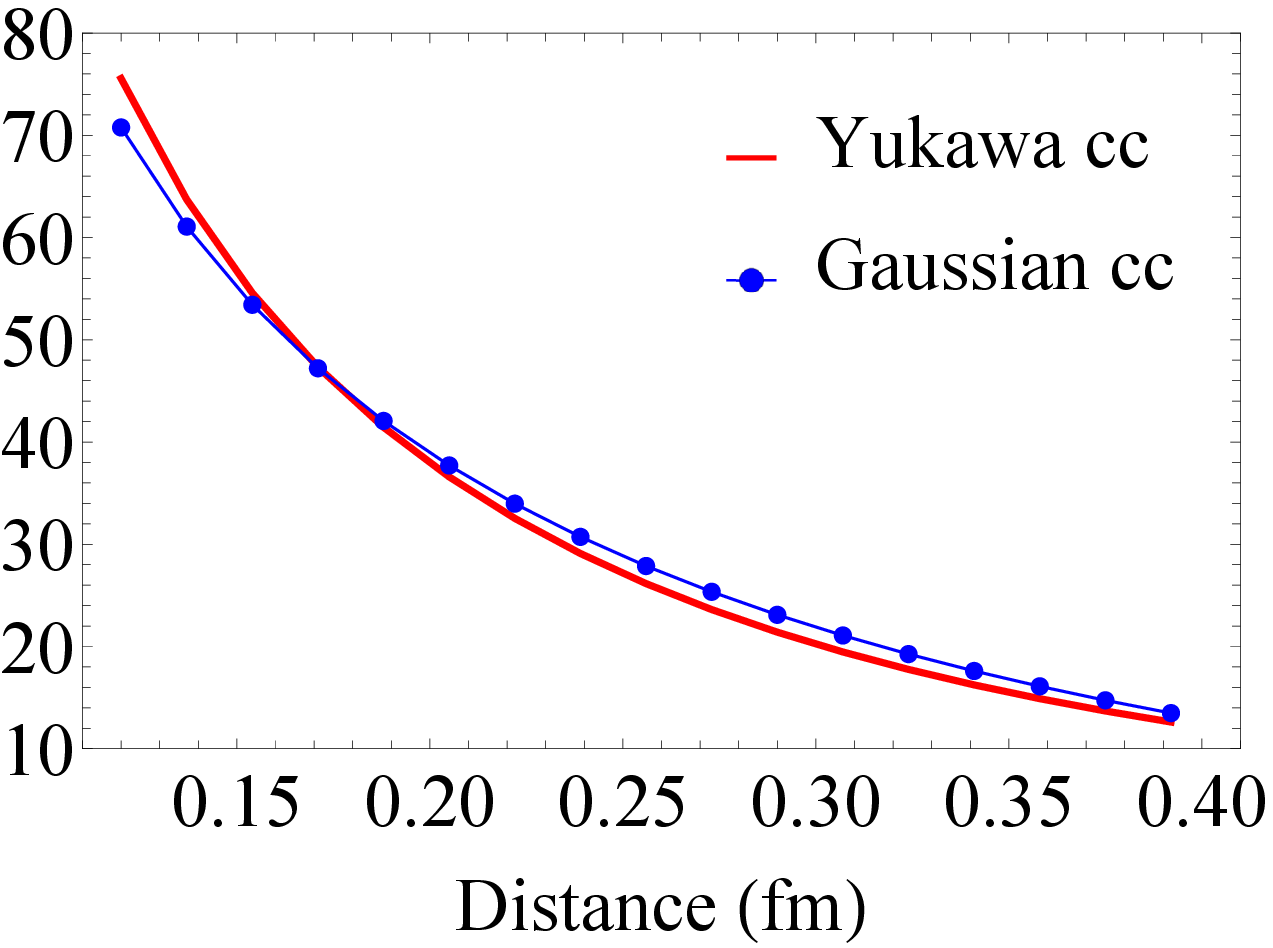}
\caption{Spatial forms of hyperfine potentials for the $\bar{u}c$ and $cc$ quark pairs. The red and blue lines represent that of the hyperfine potential used in this work and in Ref.~\cite{Noh:Prd2021}, respectively. The Gaussian type is slightly stronger in the range of 0.25 - 0.6 fm for the $\bar{u}c$ pair and 0.15 - 0.5 fm for the $cc$ pair.}
\label{ucHypPlot}
\end{figure}

\begin{figure}[h!]
\centering

\includegraphics[width=0.49\columnwidth]{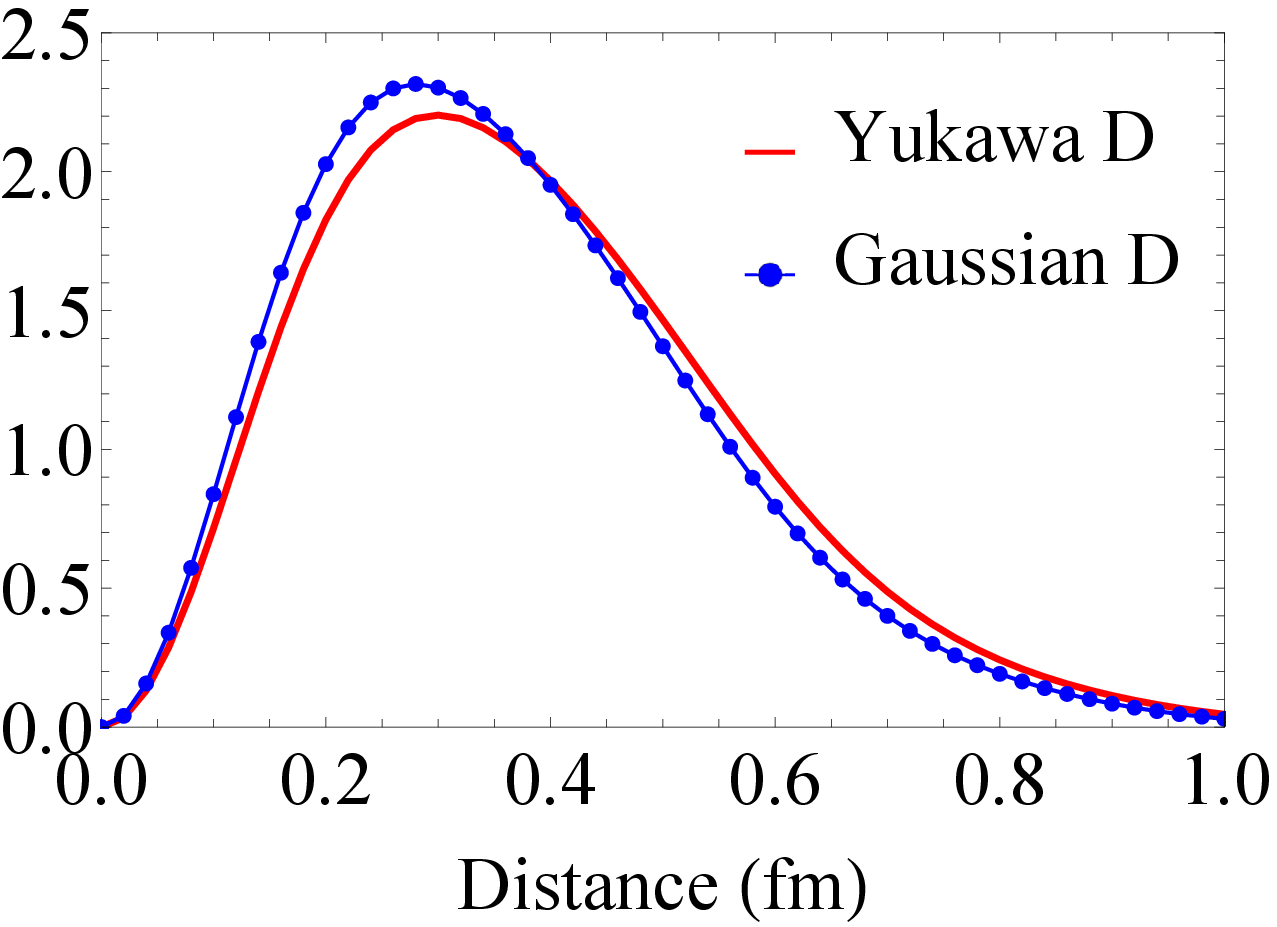}
\includegraphics[width=0.48\columnwidth]{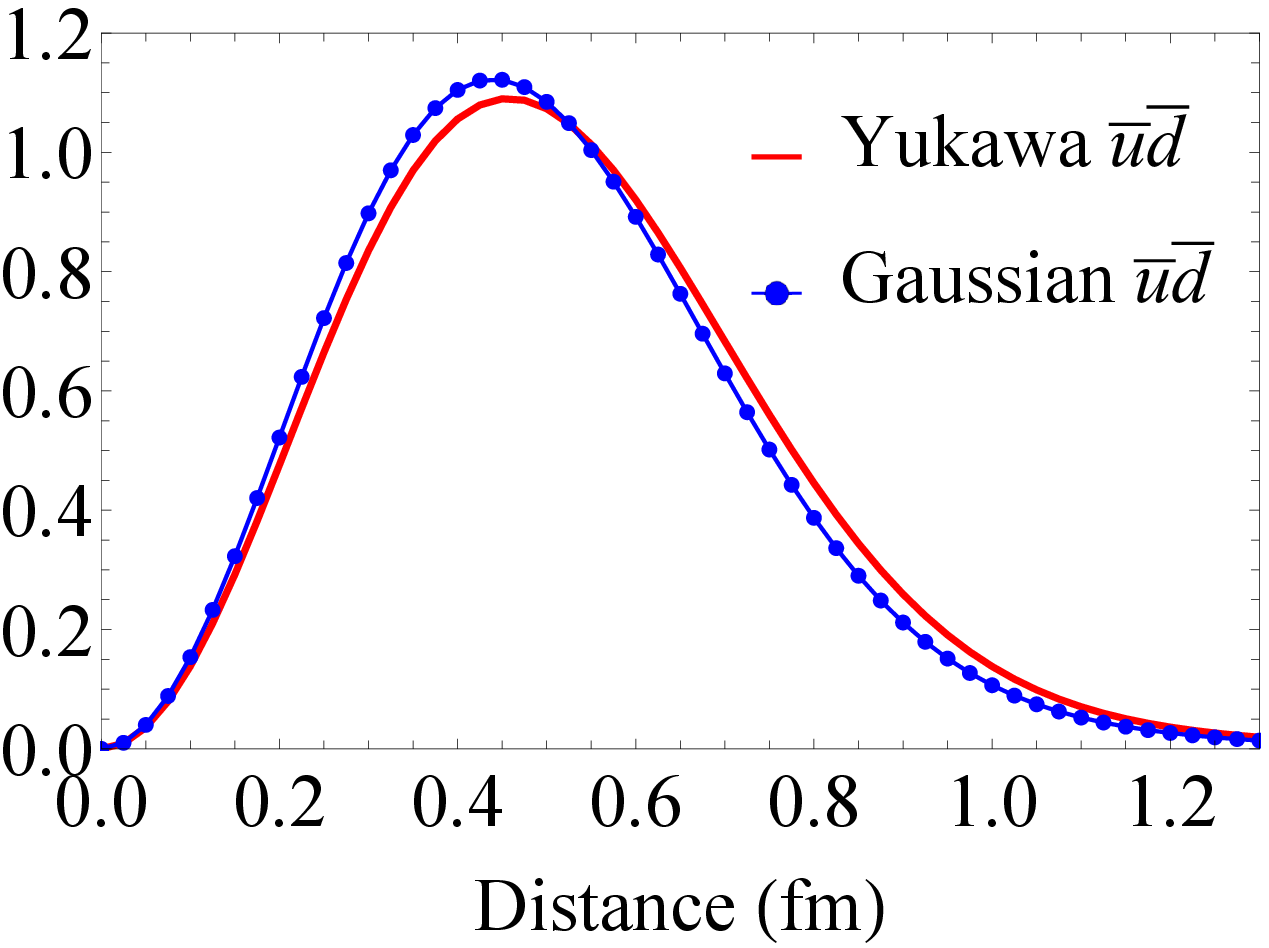}
\caption{Left panel shows the radial distribution of the probability density of the $D$ meson. Right panel shows the distribution for the $\bar{u}\bar{d}$ pair in the $T_{cc}$ state, in terms of the most dominant color $\mathbf{3}_{\bar{u}\bar{d}} \otimes \mathbf{\bar{3}}_{cc}$ state. The red and blue lines represent the distribution obtained from the hyperfine potential used in this work and in Ref.\cite{Noh:Prd2021}, respectively.}
\label{ProbDensity}
\end{figure}

We first analyze the contribution from the hyperfine potential of the $D$ or $D^*$ meson. Figure~\ref{ucHypPlot} compares the spatial functional form of the Yukawa and Gaussian type potentials, and indicates that the Yukawa potential is stronger than the Gaussian type in the vicinity of the origin. However, as shown in Table~\ref{HypSep}, the contributions from the hyperfine potential for the threshold are the opposite of what one might expect. This is due to two factors. First, for both types of potential, the peak of the probability density of the $D$ meson is at around 0.3 fm, as shown in Figure~\ref{ProbDensity}. The probability density of the $D^*$ shows almost the same as that of the $D$ and the peak is at around 0.35 fm. Furthermore, the peak value of the Gaussian type is higher than that of the Yukawa type. Second, in this region, the spatial functional form of the Gaussian type is stronger than that of the Yukawa type, as shown in Figure~\ref{ucHypPlot}.

\begin{figure}[t]
\centering

\includegraphics[width=0.49\columnwidth]{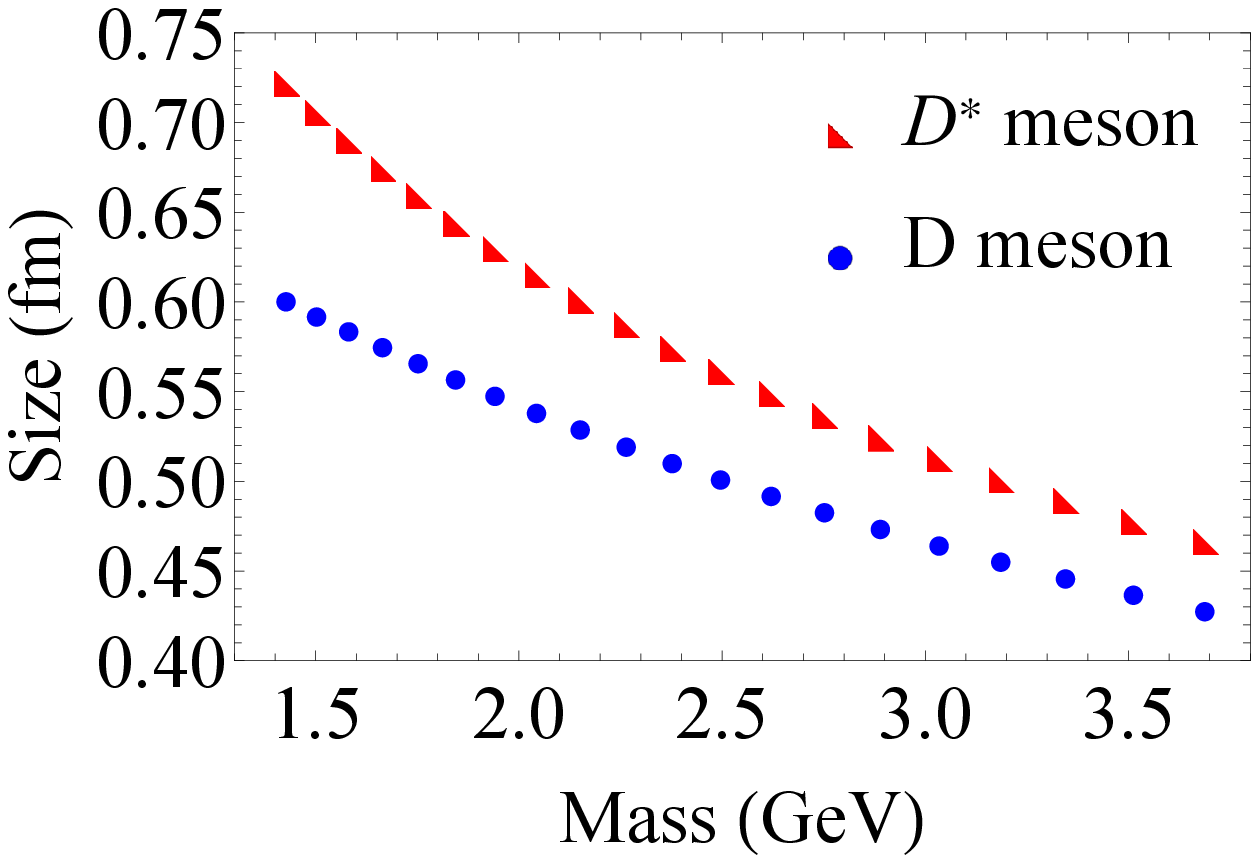}
\includegraphics[width=0.49\columnwidth]{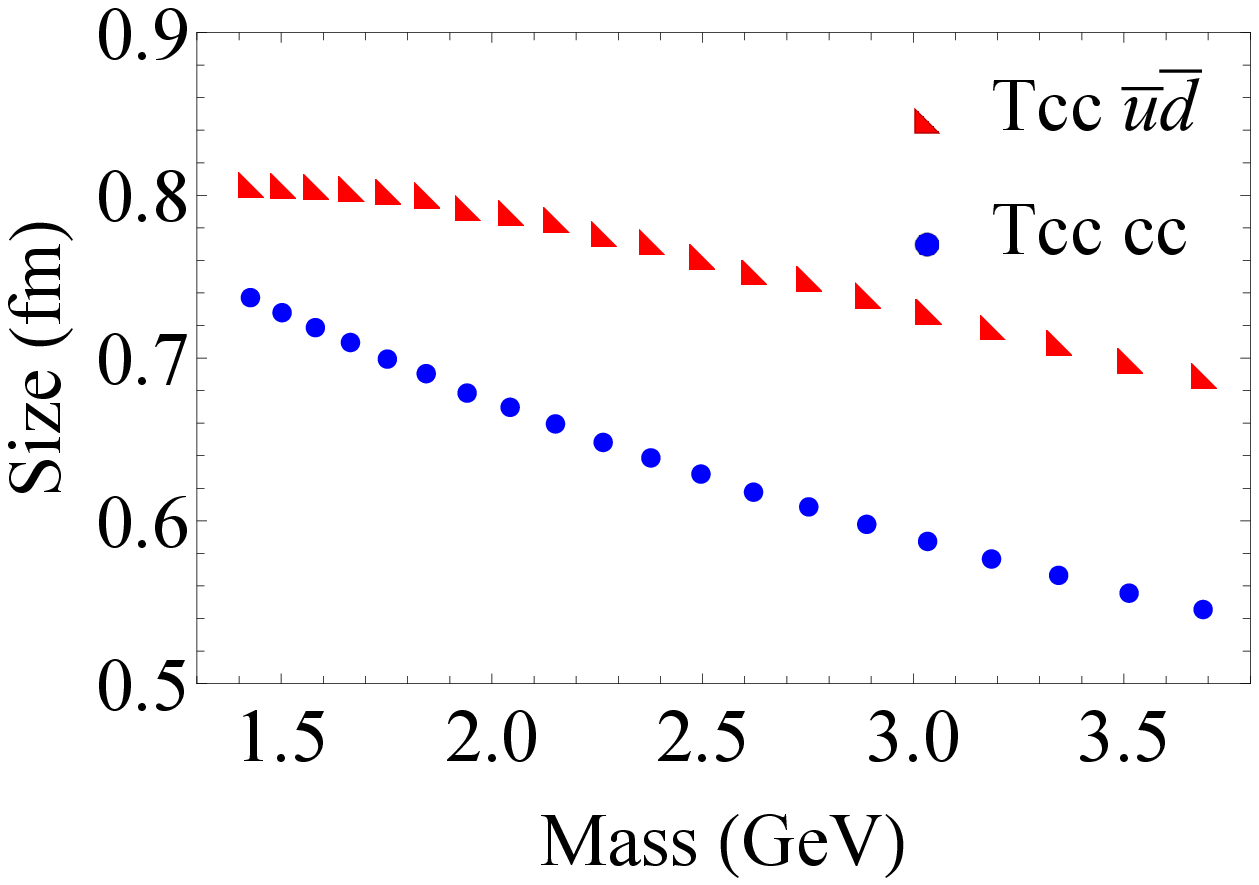}
\caption{Left panel shows the change in size of the $D$ meson with increasing constituent quark masses using spatial bases up to the 5th quanta. Right panel shows the same for the $\bar{u}\bar{d}$ and $cc$ pairs in $T_{cc}$, but using up to the 3rd quanta. The horizontal axis represents the sum of masses of the $u$ and $c$ quarks. The Gaussian type of hyperfine potential from Ref.~\cite{Noh:Prd2021} is used to obtain the figure. The notion of quanta can be referred to Ref.~\cite{Noh:Prd2021}.}
\label{SizeMassRel}
\end{figure}

Besides, Figure~\ref{SizeMassRel} shows that the size of the $D$ or $D^*$ meson is inversely proportional to the constituent quark masses. If the constituent quark masses in the Gaussian type are fitted to be lower than those used in Ref.~\cite{Noh:Prd2021}, this leads to a larger size for the $D$ or $D^*$ meson in the Gaussian type. For a proper analysis, we scale the horizontal axis with the same rate of mass change for each constituent quark. It is also possible to understand a similar behavior of $T_{cc}$ through the dependence of the relative size of $T_{cc}$ on the constituent quark masses, as shown in Figure~\ref{SizeMassRel}.

We now analyze the contributions from the hyperfine potential for each quark pair in $T_{cc}$ given in Table~\ref{HypSep}. For the $\bar{u}\bar{d}$ pair in $T_{cc}$, Figure~\ref{ProbDensity} shows that the peak of the probability density for both types of potential is located at around 0.45 fm. For the $\bar{u}c$ and $cc$ pairs, the peaks are slightly shifted from that of the $\bar{u}\bar{d}$ pair towards the origin: the peak for the $\bar{u}c$ pair ($cc$ pair) is at around 0.35 fm (0.3 fm).

For the same reason as in the case of $D$ or $D^*$, the contribution from the Gaussian type potential is found to be stronger than that of the Yukawa type for the $\bar{u}c$ and $cc$ pairs as shown in Table~\ref{HypSep}. However, for the $\bar{u}\bar{d}$ pair, the strength of the Yukawa type hyperfine potential is above that of the Gaussian type in all ranges. Thus, for the $\bar{u}\bar{d}$ pair, we find that the relative contribution from the Gaussian and Yukawa types is opposite to those of the other pairs as shown in Table~\ref{HypSep}. Furthermore, the attraction from each type of hyperfine potential mainly comes from the $\bar{u}\bar{d}$ pair. Therefore, we find that the binding energy in Eq.~(\ref{HypBinding}) obtained from the Yukawa type potential is relatively attractive as shown in Table~\ref{HypSep}. These suggest that the $\bar{u}\bar{d}$ pair plays a crucial role in the hyperfine interaction in $T_{cc}$.

To get a better understanding of the size of $T_{cc}$, we examine the probability density. In Figure~\ref{ProbDensity}, we find that the peak for the $\bar{u}\bar{d}$ pair of the Gaussian type is higher and closer to the origin than that of the Yukawa type. This trend is also observed for the $\bar{u}c$ and $cc$ pairs. Therefore, all relative quark pair sizes in $T_{cc}$ for the Gaussian type are smaller than those of the Yukawa type, as shown in Table~\ref{HypSep}.

On the other hand, we calculate the root mean square (RMS) ratio using Eq.~(10) from Ref.~\cite{Vijande:2009kj}. As discussed in Ref.~\cite{Vijande:2009kj}, a RMS ratio smaller than 1 would represent a compact configuration when the state is bound. The RMS ratio of the Yukawa type in Table~\ref{HypSep} indicates that $T_{cc}$ is highly likely to be in a compact configuration.

%%%%%%%%%%%%%%%%%%%%%%%%%%%%%%%%%%%%%%%%%%%%%%%%%%%%%%%%%%%%%%%%%%%%%%%%%%%%%%%%%%%%%%%%%%%%%%%
{\it Principal differences between $T_{cc}$ and $T_{bb}$:}
%%%%%%%%%%%%%%%%%%%%%%%%%%%%%%%%%%%%%%%%%%%%%%%%%%%%%%%%%%%%%%%%%%%%%%%%%%%%%%%%%%%%%%%%%%%%%%%
One of the most interesting findings is that the confinement potential of the $\bar{u}\bar{d}$ with $I=0$ and $cc$(or $bb$) pairs significantly contributes to the binding energy of the tetraquark. In particular, investigating the matrix element of $- \lambda^c_{i} \lambda^c_{j}$ is important because it affects the strength of the confinement potential. For both the light and heavy quark($Q$) pairs, the value with respect to the most dominant color state, $\mathbf{3}_{\bar{u}\bar{d}} \otimes \mathbf{\bar{3}}_{QQ}$, is $\frac{8}{3}$. Thus, the linearizing potential gives repulsion while the Coulomb potential gives attraction in the confinement for both pairs.

%%%%%%%%%%%%%%%%		Table 4. potentials for each pair in TQQ 		%%%%%%%%%%%%%%%%%%%%%
\begin{table}[t]

\caption{The confinement $V^{C}$ potential(in MeV) and the relative lengths $l$(in fm) for the $\bar{u}\bar{d}$ and $cc$ pairs in $T_{QQ}$. The results are obtained using the Yukawa type potential in this work.}	

\centering

\begin{tabular}{cccccc}
\hline
\hline
		Type			&	\multicolumn{2}{c}{$T_{cc}$}	&&	\multicolumn{2}{c}{$T_{bb}$}		\\
\cline{2-3}\cline{5-6}
&$\bar{u}\bar{d}$		&$cc$		&&$\bar{u}\bar{d}$	&$bb$	\\
\hline 
$V^{C} (l)$		&120(0.87)	&48(0.64)	&&	221(0.71)		&-95(0.29)	\\
\hline
\hline
\end{tabular}
\label{PotentialSep}
\end{table}
%%%%%%%%%%%%%%%%%%%%%%%%%%%%%%%%%%%%%%%%%%%%%%%%%%%%%%%%%%%%%%%%%%%%%%%%%%%%%%%%%%%%%%%%%%%%%%%%%%

For the light quark pair, the dominant part of the confinement potential comes from the linearizing potential in both $T_{cc}$ and $T_{bb}$ because their sizes are comparable to those of hadrons. However, the confinement potential in $T_{bb}$ is considerably more repulsive than that in $T_{cc}$, as shown in Table~\ref{PotentialSep}, though the size of the light quark pair in $T_{bb}$ is shorter than that of $T_{cc}$.

\begin{figure}[t]
\centering

\includegraphics[width=0.48\columnwidth]{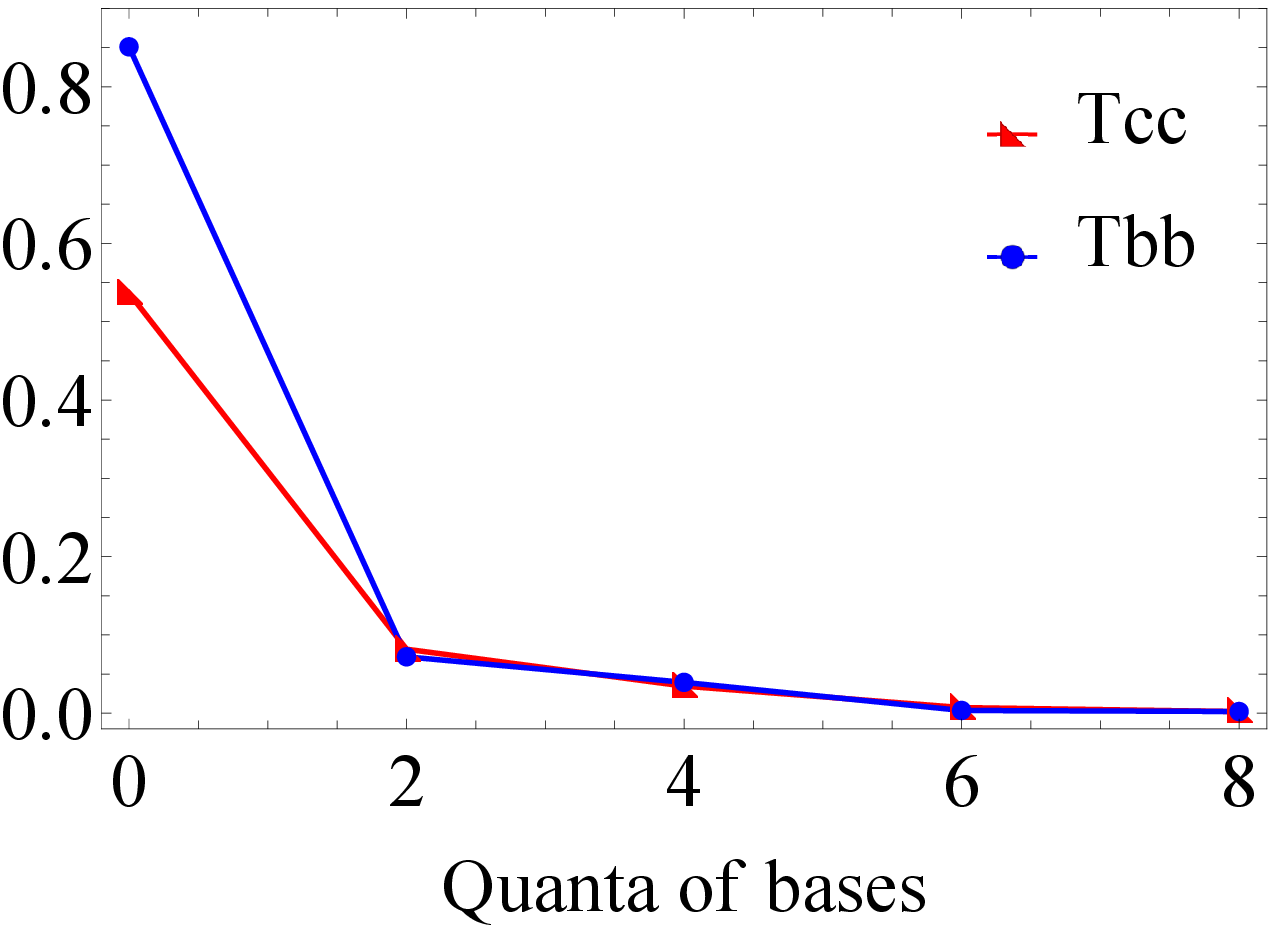}
\includegraphics[width=0.49\columnwidth]{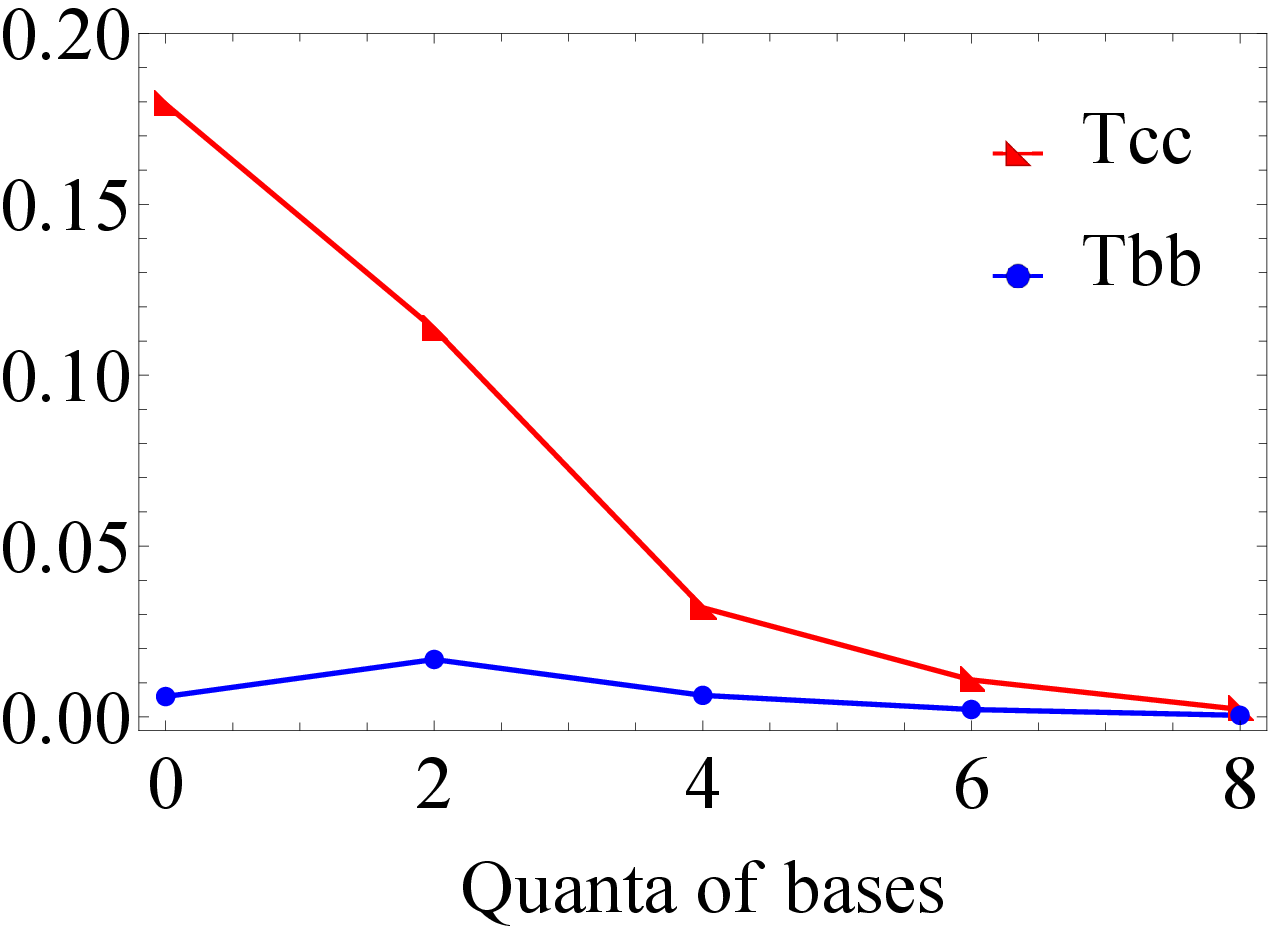}
\caption{Left(Right) panel shows the probability distribution for each quanta of spatial wave functions for $T_{cc}$ and $T_{bb}$ in terms of the color $\mathbf{3}_{\bar{u}\bar{d}} \otimes \mathbf{\bar{3}}_{QQ}$ ($\mathbf{\bar{6}}_{\bar{u}\bar{d}} \otimes {\mathbf{6}}_{QQ}$) state.}
\label{ProbAmpleQ}
\end{figure}

For the heavy quark pair, the dominant part still comes from the linearizing potential for the $T_{cc}$ despite the small size of the $cc$ pair. In contrast, for $T_{bb}$, it comes from the Coulomb potential because the size of heavy quark pair in $T_{bb}$ shrinks much shorter than that in $T_{cc}$. Thus, the confinement potential for $T_{cc}$ gives small repulsion while it gives significant attraction for $T_{bb}$. In addition to this, there is also hyperfine attraction for $T_{bb}$, which can be evaluated by Eq.~(\ref{HypBinding}) and is comparable to that of $T_{cc}$. Therefore, the ground state of $T_{bb}$ is deeply bound as shown in Table~\ref{result}.

To understand the confinement contributions in Table~\ref{PotentialSep}, it is necessary to divide the probability distribution of the ground state in terms of each color state of the tetraquark as in Figure~\ref{ProbAmpleQ}. The values of $- \lambda^c_{i} \lambda^c_{j}$ in the confinement are changed by the probability distribution for the two color states when considering the whole of bases to calculate the Hamiltonian. In Figure~\ref{ProbAmpleQ}, the contribution from the color $\mathbf{\bar{6}}_{\bar{u}\bar{d}} \otimes {\mathbf{6}}_{QQ}$ state, where the matrix element for both the light and heavy quark pairs is $-\frac{4}{3}$, is negligible for $T_{bb}$ but crucial for $T_{cc}$.
Apart from the contribution of $\mathbf{3}_{\bar{u}\bar{d}} \otimes \mathbf{\bar{3}}_{QQ}$ to the values of $- \lambda^c_{i} \lambda^c_{j}$, the contribution of $\mathbf{\bar{6}}_{\bar{u}\bar{d}} \otimes {\mathbf{6}}_{QQ}$ to $T_{cc}$ is relatively larger than that to $T_{bb}$. Therefore, the confinement potential involving the value of the matrix element for the $\bar{u}\bar{d}$ pair in $T_{cc}$ decreases much compared to that in $T_{bb}$.

Finally, we note that the significant contribution of $\mathbf{\bar{6}}_{\bar{u}\bar{d}} \otimes {\mathbf{6}}_{QQ}$ to the confinement in $T_{cc}$ is an essential feature that only appears when all the wave functions are expanded by the complete set of harmonic oscillator bases. This effect does not appear in a simple quark model~\cite{Woosung:NPA2019}, which performs calculations using a single spatial basis.

{\it Acknowledgement}
This work was supported by and by the Korea National Research Foundation under the grant number 2021R1A2C1009486(NRF).

\end{document}